\documentclass[12pt]{article}
\usepackage{amssymb}

\newcommand{\R}{\mbox{$\mathbb R$}}

\title{
${\cal PT}$-symmetric quartic anharmonic oscillator and position-dependent
mass in a perturbative approach}
\author{B. Bagchi$^a$, A. Banerjee$^a$, C. Quesne$^{b}$\\ 
{\small
$^a$ Department of Applied Mathematics, University of Calcutta,} \\ {\small 92 Acharya
Prafulla Chandra Road, Kolkata 700 009, India}\\ 
{\small $^b$ Physique Nucl\'eaire
Th\'eorique et Physique Math\'ematique,  Universit\'e Libre de Bruxelles,} \\ 
{\small Campus de la Plaine CP229, Boulevard~du Triomphe, B-1050
Brussels, Belgium}}
\date{ }
\begin{document}
\baselineskip=22pt plus 1pt minus 1pt
%%%%%%%%%%%%%%%%%%%%%%%%%%%%%%%%%%%%%%%%%%%%%%%%%%%%%%%%%%
\maketitle

\begin{abstract} 
To lowest order of perturbation theory we show that an equivalence can be
established between a $\cal PT$-symmetric generalized quartic anharmonic oscillator model
and a Hermitian position-dependent mass Hamiltonian $h$. An important feature of $h$ is
that it reveals a domain of couplings where the quartic potential could be attractive, vanishing
or repulsive. We also determine the associated physical quantities.
\end{abstract}
%
%========================================================================
%
\newpage
The interplay between pseudo-Hermitian $\cal PT$-symmetric Hamiltonians and their
equivalent Hermitian representation is currently a matter of active research \cite{mosta05,
mosta04, jones, ghosh, bagchi05}. While several case studies already exist in the literature,
Mostafazadeh, in particular, has also considered \cite{mosta05} the related issue of transition
to the classical limit. Quite significantly, he has observed that the underlying classical
Hamiltonian for the $PT$-symmetric cubic anharmonic oscillator (PTCAO), in presence of a
harmonic term, reveals the characteristics of a point particle that is endowed with a
position-dependent mass (PDM) interacting with a quartic anharmonic field.\par
%
%------------------------------------------------------------------------------
%
Motivated by Mostafazadeh's work, we recently developed \cite{bagchi06} an algorithm that
affords a categorization of a whole class of quantal PDM Hamiltonians in terms of
perturbatively equivalent $\cal PT$-symmetric counterparts with configuration space
\R. The PTCAO is one such model, which was shown to have a Hermitian PDM partner
Hamiltonian.\par
%
%------------------------------------------------------------------------------
%  
Keeping $\cal PT$ symmetry intact, the PTCAO scheme can obviously be extended  to include the one-parameter family of potentials $x^2 ({\rm i}
x)^{\delta}$, $\delta$ real. It is curious to note that $\delta=2$ implies quartic anharmonicity
but with a wrong sign (see, e.g., \cite{bender05}). However, as Bender and Boettcher
argued \cite{bender98} (see also \cite{znojil}), such a feature facilitates its quasi-exact
solvability.\par
%
%----------------------------------------------------------------------------
%
In this paper, we take up the study of a generalized scheme \cite{banerjee, bender06} of
$PT$-symmetric quartic anharmonic oscillator (PTQAO) in the spirit of the perturbative
analysis of \cite{bagchi06} and examine when it should map to a Hermitian PDM Hamiltonian.
We also estimate the fourth-order contribution to it and write down the physical position and
momentum operators up to third order. Finally, we determine the classical limit.\par
%
%=============================================
%
Consider the harmonic oscillator perturbed by an imaginary cubic term to the first order and a
$\delta=2$ quartic anharmonic term to the second order as follows:
\begin{equation}
  {\cal H} = {\cal H}_0 + \varepsilon {\cal H}_1 + \varepsilon^2 {\cal H}_2. \label{eq:H}
\end{equation}
The various terms in the Hamiltonian $\cal H$ are
\begin{eqnarray}
  {\cal H}_0 & = & \frac{p^2}{2\mu_0} + a v_1^{(\rm r)}(x), \qquad v_1^{(\rm r)}(x) =
       x^2, \nonumber \\ 
  {\cal H}_1 & = & {\rm i} b v^{(\rm i)}(x), \qquad v^{(\rm i)}(x) = x^3, \nonumber \\
  {\cal H}_2 & = & - c v_2^{(\rm r)}(x), \qquad v_2^{(\rm r)}(x) = x^4, 
\end{eqnarray}
with $\varepsilon \ll 1$, $a$, $b$ ($\ne 0$) and $c$ are positive coupling constants,
$v_j^{(\rm r)}(-x) = v_j^{(\rm r)}(x) \in \R$, $j=1$, 2, $v^{(\rm i)}(-x) = - v^{(\rm i)}(x) \in
\R$.\par
%
%-------------------------------------------------------------------------------
%
Resorting to dimensionless variables proves convenient and requires the following
transformations to be enforced:
\begin{eqnarray}
  X & = & \ell^{-1} x, \qquad P = \ell \hbar^{-1} p, \nonumber \\
  \alpha & = & \ell^4 \hbar^{-2} \mu_0 a, \qquad \beta = \ell^3 \hbar^{-2} \mu_0 b,
       \qquad \gamma = \ell^2 \hbar^{-2} \mu_0 c, \qquad \epsilon = \ell^2 \varepsilon, 
\end{eqnarray}
where $\ell$ is the length scale.\par
%
%--------------------------------------------------------------------------------
%
The Hamiltonian $\cal H$ then gets modified to
\begin{equation}
  H = \nu^{-1} {\cal H} = H_0 + \epsilon H_1 + \epsilon^2 H_2, \qquad \nu = \frac{\hbar^2}
  {\mu_0 \ell^2}, \label{eq:Hbis}
\end{equation}
with
\begin{equation}
  H_0 = \frac{P^2}{2} + V_1^{(\rm r)}(X), \qquad H_1 = {\rm i} V^{(\rm i)}(X), \qquad
  H_2 = - V_2^{(\rm i)}(X), \label{eq:Hter}
\end{equation}
where $V_1^{(\rm r)} = \alpha X^2$, $V^{(\rm i)} = \beta X^3$ and $V_2^{(\rm r)} =
\gamma X^4$.\par
%
%----------------------------------------------------------------------------------
% 
As in \cite{bagchi06}, we can set up, corresponding to (\ref{eq:H}), an equivalent Hermitian
Hamiltonian $h(x, p)$ which reduces to some PDM Hamiltonian to lowest order in
$\varepsilon$. Noting that $h(X, P) = \nu^{-1} h(x, p)$, we therefore write
\begin{equation}
  h(X, P) = H_0(X, P) + \epsilon^2 h^{(2)}(X, P) + \epsilon^4 h^{(4)}(X, P) + O(\epsilon^6),
  \label{eq:h}
\end{equation}
where $H_0(X, P) + \epsilon^2 h^{(2)}(X, P) = P \frac{1}{2m(X)} P + V_{\rm eff}(X)$,
$1/m(X) = 1 + \epsilon^2 M^{(2)}(X)$, and $V_{\rm eff}(X) = V_1^{(\rm r)}(X) +
\epsilon^2 V_{\rm eff}^{(2)}(X)$. Comparing terms of $O(\epsilon^2)$ produces
\begin{equation}
  h^{(2)}(X, P) = \frac{1}{2} P M^{(2)}(X) P + V_{\rm eff}^{(2)}(X). \label{eq:h2}   
\end{equation}
\par
%
%--------------------------------------------------------------------------------------
% 
In a pseudo-Hermitian theory, for the spectrum of a diagonalizable Hamiltonian to be real, it is
necessary \cite{mosta04} that such an operator be Hermitian with respect to a
positive-definite inner product $\langle \cdot, \cdot \rangle_+$. The latter may be expressed
in terms of the defining inner product $\langle \cdot, \cdot \rangle$ as
\begin{equation}
  \langle \cdot, \cdot \rangle_+ = \langle \cdot, \eta_+ \cdot \rangle, \label{eq:inner}
\end{equation}
where the positive-definite metric operator $\eta_+: {\cal L} \to {\cal L}$ of the reference
Hilbert space $\cal L$, in which the Hamiltonian acts, belongs to the set of all Hermitian
invertible operators. The Hilbert space equipped with the inner product (\ref{eq:inner}) may be identified as the
physical Hilbert space ${\cal L}_{\rm phys}$.\par
%
%--------------------------------------------------------------------------------------
% 
Pseudo-Hermiticity of Hamiltonian $\cal H$ with respect to $\eta_+$ is defined by
\begin{equation}
  {\cal H}^{\dagger} = \eta_+ {\cal H} \eta_+^{-1} \label{eq:pseudo}
\end{equation}
and serves as a necessary and sufficient condition for $\cal H$ to possess a real
spectrum. Furthermore, given $\cal H$, the equivalent Hermitian operator would be
\begin{equation}
  h(x, p) = \rho {\cal H} \rho^{-1}, \label{eq:hbis}
\end{equation}
with $\rho = \sqrt{\eta_+}$.\par
%
%--------------------------------------------------------------------------------------
%
The metric $\eta_+$ can be represented by
\begin{equation}
  \eta_+ = e^{- Q(X, P)}, \qquad Q(X, P) = \epsilon Q_1(X, P) + \epsilon^3 Q_3(X, P) +
  \cdots, \label{eq:eta}
\end{equation}
where every $Q_j(X, P)$, $j=1$, 3,~\ldots, is Hermitian, symmetric in $X$ and antisymmetric
in $P$. Using (\ref{eq:eta}), the following relations emerge when equation (\ref{eq:Hbis})
along with (\ref{eq:Hter}) are substituted in the dimensionless counterpart of
(\ref{eq:pseudo}):
\begin{eqnarray}
  \left[\frac{P^2}{2} + V_1^{(\rm r)}, Q_1\right] & = & - 2{\rm i} V^{(\rm i)},
        \label{eq:cond1} \\
  \left[\frac{P^2}{2} + V_1^{(\rm r)}, Q_3\right] & = & - \frac{1}{6} \left[Q_1, \left[Q_1, 
       {\rm i} V^{(\rm i)}\right]\right] - \left[Q_1, V_2^{(\rm r)}\right]. \label{eq:cond2} 
\end{eqnarray}
\par
%
%---------------------------------------------------------------------------------
% 
With $h(X, P)$ prescribed by (\ref{eq:h}), it is easy to deduce from the dimensionless
counterpart of (\ref{eq:hbis}) and the Baker-Campbell-Hausdorff identity that
\begin{eqnarray}
  h^{(2)}(X, P) & = & H_2 + \frac{1}{4} [H_1, Q_1], \label{eq:h2bis} \\
  h^{(4)}(X, P) & = & \frac{1}{4} [H_1, Q_3] - \frac{1}{192} [[[H_1, Q_1], Q_1], Q_1].
       \label{eq:h4}
\end{eqnarray}
Comparison of (\ref{eq:h2bis}) with (\ref{eq:h2}) then yields
\begin{equation}
  \frac{1}{2} P M^{(2)} P + V_{\rm eff}^{(2)} = - V_2^{(\rm r)} + \frac{1}{4} \left[{\rm i}
  V^{(\rm i)}, Q_1\right]. \label{eq:h2-h2bis}
\end{equation}
\par
%
%------------------------------------------------------------------------
%
In Ref.~\cite{bagchi06}, it was noted that the operator $Q_1$ admits of a normal form
\begin{equation}
  Q_1 = - {\rm i} \sum_{k=0}^{\infty} S_k(X) \frac{d^k}{dX^k}, \label{eq:Q1}  
\end{equation}
where
\begin{eqnarray}
  && S_{2k} = \sum_{j=k}^{\infty} (-1)^j
       \left(\begin{array}{c}
          2j+1 \\
          2k
       \end{array}\right) 
       \frac{d^{2j-2k+1} R_j}{dX^{2j-2k+1}}, \nonumber \\ 
  && S_{2k+1} = \sum_{j=k}^{\infty} (-1)^j (1 + \delta_{j,k})  
        \left(\begin{array}{c}
        2j+1 \\
        2k+1
     \end{array}\right)
     \frac{d^{2j-2k} R_j}{dX^{2j-2k}}, 
\end{eqnarray}
and $R_j$'s are appropriate even functions of $X$. A similar expansion can be carried out for
the operator $Q_3$.\par
%
%--------------------------------------------------------------------------------
%
Consider $Q_1$ first. Substituting (\ref{eq:Q1}) in equation (\ref{eq:cond1}), we find after a
little algebra
\begin{eqnarray}
  && \frac{1}{2} \frac{d^2 S_0}{dX^2} + \sum_{j=1}^{\infty} S_j \frac{d^j V_1^{(\rm r)}}
        {dX^j} = - 2 V^{(\rm i)}, \nonumber \\
  && \frac{1}{2} \frac{d^2 S_k}{dX^2} +\frac{dS_{k-1}}{dX} +  \sum_{j=k+1}^{\infty} 
        \left(\begin{array}{c}
          j \\
          k
        \end{array}\right)
        S_j \frac{d^{j-k}V_1^{(\rm r)}}{dX^{j-k}} = 0, \qquad k=1, 2, \ldots.
\end{eqnarray}
\par
%
%-------------------------------------------------------------------------
%
On the other hand, equation (\ref{eq:h2-h2bis}) leads to
\begin{equation}
  -4 \left(V^{(2)}_{\rm eff} + V_2^{(\rm r)}\right) = W_0, \quad 2 \frac{dM^{(2)}}{dX} =
  W_1, \quad 2 M^{(2)} = W_2, \quad W_k = 0 \quad k=3, 4, \ldots, 
\end{equation}
with
\begin{equation}
  W_k \equiv \sum_{j=k+1}^{\infty}
        \left(\begin{array}{c}
          j \\
          k
        \end{array}\right)
        S_j \frac{d^{j-k} V^{(\rm i)}}{dX^{j-k}}. 
\end{equation}
\par
%
%---------------------------------------------------------------------------------------
%
{}For the problem of PTQAO at hand, it is evident that $Q_1$ contains up to cubic power in
$P$ only, so that the sum over $k$ in (\ref{eq:Q1}) is restricted from 0 to 3. This implies
\cite{bagchi06} $R_k = 0$, $k=2$, 3,~\ldots, and $S_k = 0$, $k=4$, 5,~\ldots. Our results,
corresponding to $V_1^{(\rm r)}$, $V^{(\rm i)}$ and $V_2^{(\rm r)}$ given by
(\ref{eq:Hter}), are summarized below:
\begin{eqnarray}
  S_0 & = & - \frac{\beta}{\alpha} X, \qquad S_1 = - \frac{\beta}{\alpha} X^2, \qquad
       S_3 = \frac{\beta}{3\alpha^2}, \nonumber \\
  V^{(2)}_{\rm eff} & = & \frac{1}{4\alpha} (3 \beta^2 - 4 \alpha \gamma) X^4 -  
       \frac{\beta^2}{2\alpha^2}, \qquad M^{(2)} = \frac{3\beta^2}{2\alpha^2} X^2. 
\end{eqnarray}
\par
%
%--------------------------------------------------------------------------------
% 
On the other hand, the operator $Q_3$ contains up to fifth power in $P$ only. The coefficient
functions can be calculated by solving equation (\ref{eq:cond2}) and the results used to
determine $h^{(4)}(X, P)$ through equation (\ref{eq:h4}).\par
%
%---------------------------------------------------------------------------------------
%
{}From (\ref{eq:h}) we find, on going back to variables and operators with dimensions, the
following form of the equivalent Hermitian PDM Hamiltonian to PTQAO:
\begin{eqnarray}
  h(x, p) & = & \frac{1}{2\mu_0} p \left(1 + \frac{3 \varepsilon^2 b^2}{2 a^2} x^2\right) p
        + a x^2 - \frac{\hbar^2 \varepsilon^2 b^2}{2\mu_0 a^2} + \frac{\varepsilon^2}{4a}
        (3 b^2 - 4ac) x^4 \nonumber \\
  && \mbox{} + \varepsilon^4 h^{(4)}(x, p) + O(\varepsilon^6), 
\end{eqnarray}
where
\begin{eqnarray}
  h^{(4)}(x, p) & = & \frac{b^4}{32 a^6} \biggl(\frac{p^6}{\mu_0^3} - \frac{18
        a}{\mu_0^2} \{x^2, p^4\} - \frac{51 a^2}{2 \mu_0} \{x^4, p^2\} - 14 a^3 x^6 -
        \frac{81 \hbar^2 a}{\mu_0^2} p^2 \nonumber \\
  && \mbox{} - \frac{138 \hbar^2 a^2}{\mu_0} x^2\biggr) + \frac{3 b^2 c}{2 a^4}
        \biggl(\frac{1}{2 \mu_0^2} \{x^2, p^4\} + \frac{3 a}{2 \mu_0} \{x^4, p^2\} +
        a^2 x^6 \nonumber \\
  && + \mbox{} \frac{2 \hbar^2}{\mu_0^2} p^2 + \frac{8 \hbar^2 a}{\mu_0} x^2\biggr)
\end{eqnarray}
has been written in terms of anticommutators. It is evident that inclusion of higher-order
corrections will make the structure of $h(x, p)$ more complicated.\par
%
%--------------------------------------------------------------------------------------
%
Using the similarity transformation induced by $\rho$, it is straigthforward to obtain the
physical position and momentum operators \cite{mosta04}
\begin{eqnarray}
  x_{\rm phys} & = & x + \frac{\rm i \varepsilon b}{2 \mu_0 a^2} (p^2 + \mu_0 a x^2) +
       \frac{\varepsilon^2 b^2}{8 \mu_0 a^3}(\{x, p^2\} - 2 \mu_0 a x^3) \nonumber \\
  && \mbox{} - \frac{{\rm i}\varepsilon^3 b^3}{8 \mu_0^2 a^5} [5 p^4 + 6\mu_0 a \{x^2,
       p^2\} + \mu_0 a (5\mu_0 a x^4 + 3 \hbar^2)] \nonumber \\
  &&  \mbox{} + \frac{{\rm i}\varepsilon^3 bc}{2 \mu_0^2 a^4} [2 p^4 + 3\mu_0 a \{x^2,
       p^2\} + 2\mu_0 a (\mu_0 a x^4 + \hbar^2)], \\
  p_{\rm phys} & = & p - \frac{{\rm i} \varepsilon b}{2a} \{x, p\} + \frac{\varepsilon^2 b^2}
       {8 \mu_0 a^3} (2 p^3 - \mu_0 a \{x^2, p\}) \nonumber \\
  && \mbox{}  + \frac{{\rm i}\varepsilon^3 b^3}{4 \mu_0 a^4} (\{x, p^3\} + 4 \mu_0 a
       \{x^3, p\}) \nonumber \\
  && \mbox{} -  \frac{{\rm i}\varepsilon^3 bc}{\mu_0 a^3} (\{x, p^3\} + 2 \mu_0 a
       \{x^3, p\}),
\end{eqnarray} 
where the results have been written up to third order.\par
%
%---------------------------------------------------------------------------------
%
{}Finally, the classical Hamiltonian ${\cal H}_{\rm c}(x_{\rm c}, p_{\rm c})$ can be derived
by replacing $x$ and $p$ in $h(x, p)$ by the classical variables $x_{\rm c}$ and $p_{\rm
c}$ and proceeding to the limit $\hbar \to 0$ (assuming that the limit exists), i.e., ${\cal
H}_{\rm c}(x_{\rm c}, p_{\rm c}) = \lim_{\hbar \to 0} h(x_{\rm c}, p_{\rm c})$. Our result
is
\begin{eqnarray}
  {\cal H}_{\rm c} & = & \frac{p_{\rm c}^2}{2\mu_0} + a x_{\rm c}^2 + \varepsilon^2 \left[
       \frac{3 b^2}{4 a^2} \left(\frac{1}{\mu_0} x_{\rm c}^2 p_{\rm c}^2 + a x_{\rm c}^4
       \right) - c x_{\rm c}^4\right] \nonumber \\
  && \mbox{} + \varepsilon^4 \biggl[\frac{b^4}{32 a^6} \left(\frac{p_{\rm c}^6}{\mu_0^3}
       - \frac{36a}{\mu_0^2} x_{\rm c}^2 p_{\rm c}^4 - \frac{51 a^2}{\mu_0} x_{\rm c}^4
       p_{\rm c}^2 - 14 a^3 x_{\rm c}^6\right) \nonumber \\
  && \mbox{} + \frac{3 b^2 c}{2 a^4} \left(\frac{1}{\mu_0^2} x_{\rm c}^2 p_{\rm c}^4 +
       \frac{3a}{\mu_0} x_{\rm c}^4 p_{\rm c}^2 + a^2 x_{\rm c}^6\right)\biggr] + 
       O(\varepsilon^6). \label{eq:Hc}
\end{eqnarray}
It is clear from (\ref{eq:Hc}) that, for sufficiently small $\varepsilon$ so that terms of order
$\varepsilon^4$ and higher can be neglected, ${\cal H}_{\rm c}$ describes a point particle
with a PDM
\begin{equation}
  \frac{\mu_0}{1 + 3 \varepsilon^2 b^2 x_{\rm c}^2/(2a^2)} \simeq \mu_0\left(1 - \frac{3
  \varepsilon^2 b^2}{2a^2} x_{\rm c}^2\right), \label{eq:PDM}   
\end{equation}
interacting with a quartic anharmonic potential of strength $\varepsilon^2 (3b^2 -
4ac)/(4a)$. It is worth noting that the PDM (\ref{eq:PDM}) is the same as that obtained for
the PTCAO \cite{mosta05}. Also, an important point to observe is that the quartic potential
may be attractive, null or repulsive according to whether $3b^2 - 4ac$ is positive, vanishing or
negative.\par
%
%-----------------------------------------------------------------------------------
% 
We should point out that employing $x_{\rm phys}$ and $p_{\rm phys}$ would enable one
to obtain their classical counterparts. However, this would necessitate a formulation of
classical mechanics having a complex phase-space structure \cite{kaushal, mosta06}. This is
beyond the scope of the present paper.\par
%
%--------------------------------------------------------------------------------------
%
To conclude, we have explored the relationship between the PTQAO model and the
corresponding Hermitian PDM Hamiltonian in the framework of perturbation theory. We have
also constructed the nonlocal physical position and momentum operators  and the classical PDM
Hamiltonian associated with PTQAO.\par
%
%=================================================
%
\bigskip
We thank Prof.\ A.\ Mostafazadeh for an enlightening correspondence on the relevance of the
classical limit. BB gratefully acknowledges the support of the National Fund for Scientific
Research (FNRS), Belgium, and the warm hospitality at PNTPM, Universit\'e Libre de Bruxelles,
where part of this work was carried out. CQ is a Research Director of the National Fund for
Scientific Research (FNRS), Belgium.\par
%
%=============================================
%

\end{document}